\documentclass[
aip,
preprint
]{revtex4-1}
\usepackage[utf8]{inputenc}
\usepackage{graphicx}
\usepackage{siunitx}
\usepackage{xcolor}
\usepackage{textgreek}
\usepackage{hyperref}
\hypersetup{
    colorlinks,
    citecolor=blue,    filecolor=blue,
    linkcolor=blue,     urlcolor=blue
}

\begin{document}

% \title{Carrier-envelope phase-controlled laser-plasma acceleration of relativistic electron beams}
\title{Waveform control of relativistic electron dynamics in an underdense plasma}
\date{\today}

\author{Julius Huijts}
 \thanks{These authors contributed equally}
%\email{julius.huijts@ensta-paris.fr} 
\author{Lucas Rovige}
 \thanks{These authors contributed equally}
\author{Igor A. Andriyash}
\author{Aline Vernier}
\author{Marie Ouillé}
\author{Jaismeen Kaur}
\author{Zhao Cheng}
\author{Rodrigo Lopez-Martens}
\author{Jérôme Faure}
\affiliation{LOA, CNRS, Ecole Polytechnique, ENSTA Paris, Institut Polytechnique de Paris, 181 Chemin de la Hunière et des Joncherettes, 91120 Palaiseau, France}

\maketitle

% NO ABSTRACT

% Introduction of approximately 150 words, summarizing the background, rationale, main results (introduced by "Here we show" or some equivalent phrase) and implications of the study. This paragraph should be referenced, as in Nature style, and should be considered part of the main text, so that any subsequent introductory material avoids too much redundancy with the introductory paragraph.

\textbf{The interaction of ultra-intense laser pulses with an underdense plasma is used in laser-plasma acceleration\cite{tajima_laser_1979,esarey_physics_2009} to create compact sources of  ultrashort pulses of relativistic electrons and X-rays\cite{corde_femtosecond_2013}. The accelerating structure is a plasma wave, or wakefield, that is excited by the laser ponderomotive force, a force that is usually assumed to depend solely on the laser envelope and not on its exact waveform. Here, we use near-single-cycle laser pulses with a controlled carrier-envelope-phase (CEP)\cite{jones2000carrier} to show that the actual waveform of the laser field has a clear impact on the plasma response. We measure relativistic electron beams that are found to be strongly CEP dependent, implying that we achieve waveform control of electron dynamics in underdense laser-plasma interaction. Our results pave the way to high precision, sub-cycle control of electron injection in plasma accelerators, enabling the production of attosecond relativistic electron bunches and X-rays.}

%Laser-plasma acceleration\cite{tajima_laser_1979,esarey_physics_2009} relies on the interaction of ultra-intense laser pulses with an underdense plasma for producing ultrashort pulses of relativistic electrons and X-rays\cite{corde_femtosecond_2013} in a compact manner. 
% start main text
%Strong field on solids: \cite{apolonski2004observation}, nanotips \cite{kruger2011attosecond}
Waveform control has revolutionized several domains of laser-matter interaction, as it allows for an extremely high degree of control on electron dynamics. First applied in the frequency-domain to drastically improve atomic clock precision\cite{diddams2001optical}, it was quickly adopted by the strong-field physics community to exert coherent control on electron dynamics in photoionization\cite{paulus_absolute_2001}, high-harmonic generation in gases\cite{baltuska_attosecond_2003,nisoli2003effects} as well as laser-induced fragmentation of molecules \cite{kling_control_2006}.  
Waveform control in laser-plasma interaction has been more difficult to achieve because it requires few-cycle, CEP-stable laser pulses at intensities higher by several orders of magnitude. In laser-plasma interaction, CEP control was first demonstrated in overdense plasmas (i.e. solid targets) and at moderate intensity\cite{borot2012attosecond}, with the final goal of producing isolated attosecond pulse of XUV radiation\cite{wheeler2012attosecond}. Recently, several experiments on solid targets at relativistic intensity have displayed CEP effects and their potential\cite{kormin2018spectral,jahn2019towards,bohle2020generation}. In experiments on solid targets, the interaction takes place at the plasma surface where  the CEP has a fixed value, but this is no longer true in underdense plasmas as the laser can propagate in the medium. During propagation, the CEP slips because the envelope travels at the group velocity $v_g$ which differs from the laser phase velocity $v_{\phi}$. The length scale over which dispersion changes the CEP by $2\pi$ can then be estimated as\cite{nerush_carrier-envelope_2009,faure_review_2018}:
\begin{equation}
    L_{2\pi}=\frac{c}{v_{\phi}-v_g}\lambda_0,
\end{equation}
where $c$ is the velocity of light and $\lambda_0$ is the laser wavelength. Typical propagation lengths are many times larger than $L_{2\pi}$, which averages out CEP effects. As a result, CEP-controlled electron dynamics can only be observed if these dynamics are governed by a factor that is localized to a fraction of $L_{2\pi}$, so they can be associated to a specific CEP value. Additionally, a high level of control is required over not only the laser waveform, but also over the plasma parameters as they strongly influence dispersion. Therefore, except for preliminary results\cite{faure_review_2018} showing hints of CEP-dependent electron spectra in a laser-plasma accelerator (LPA), CEP effects in underdense laser-plasma interaction have remained elusive until now.

In a LPA, the laser-driven wakefield traps and accelerates electrons to relativistic energies over very short distances\cite{esarey_physics_2009}. Wakefield excitation is particularly efficient when the laser pulse length $c\tau$ is resonant with the plasma wavelength, i.e. when $c\tau\simeq\lambda_p/2$. Plasma electrons are then self-injected and accelerated into the wakefield, forming high quality, quasi-monoenergetic electron beams\cite{geddes_high-quality_2004,faure_laserplasma_2004,mangles_monoenergetic_2004}. While most LPAs are driven by laser pulses containing many optical cycles, few-cycle pulses can now be used to excite wakefields and accelerate electrons\cite{schmid_few-cycle_2009,guenot_relativistic_2017,rovige_optimization_2021,salehi2020laseraccelerated}. In this case, the usual framework of a cycle-averaged ponderomotive force\citep{mora_kinetic_1997} is not sufficient to describe the interaction. Instead, the actual waveform of the laser pulse needs to be considered, in particular the phase between the envelope and the carrier wave (the CEP). 

Theory and simulation studies show that the precise control of the laser waveform through the CEP can have a strong impact in LPAs. Simulations indicate that single-cycle laser pulses cause significant asymmetries of the plasma wakefield\cite{zhidkov_electron_2008,nerush_carrier-envelope_2009}. Nerush \emph{et al.}\cite{nerush_carrier-envelope_2009} found that these asymmetries are due to higher-order terms of the plasma response and are CEP dependent. As the laser pulse propagates in the plasma, the CEP-slippage causes the wake to oscillate transversely on the length scale of $L_{2\pi}$. As seen in recent simulations\cite{huijts2021identifying,xu_periodic_2020}, the asymmetric plasma wakefield can in turn cause the off-axis injection of sub-femtosecond electron bunches, with a collective, non-zero transverse momentum. As the electrons are accelerated, they oscillate collectively until they leave the plasma with an off-axis beam pointing that is CEP dependent. 

\begin{figure}[t!]
    \includegraphics[width=0.6\textwidth]{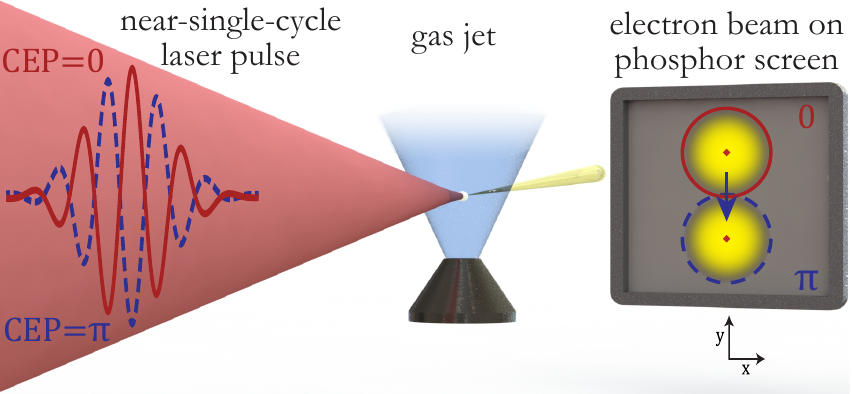}
    \caption{\label{fig:principle}\textbf{Principle of the experiment.} An intense, near-single-cycle laser pulse is focused into a gas jet, where it ionizes the gas and drives a plasma wake. Through laser wakefield acceleration, electrons are accelerated to relativistic energies. A phosphor screen is used to image the electron beam. The shape of the electric field of the laser pulse is controlled through the carrier-envelope phase (CEP). Varying the CEP in the experiment changes the pointing of the electron beam.}
\end{figure}

In our experiments, we are able to detect and control this CEP-dependent pointing of the accelerated electron beam, see illustration in figure~\ref{fig:principle}. Our LPA is driven by laser pulses as short as 1.3 cycles ($< 4$~fs FWHM in intensity)\cite{guenot_relativistic_2017,faure_review_2018,rovige_optimization_2021}. Two active feedback loops allow for stabilization and control of the CEP (see Methods). The laser pulse is focused into a supersonic nitrogen gas jet where it resonantly drives a plasma wake, in which electrons are accelerated to a few MeV. After acceleration, these electrons leave the gas jet and propagate to a set of diagnostics which measure either the electron beam charge and angular distribution, or its energy distribution using a magnetic electron spectrometer.

\begin{figure*}
    \includegraphics[width=\textwidth]{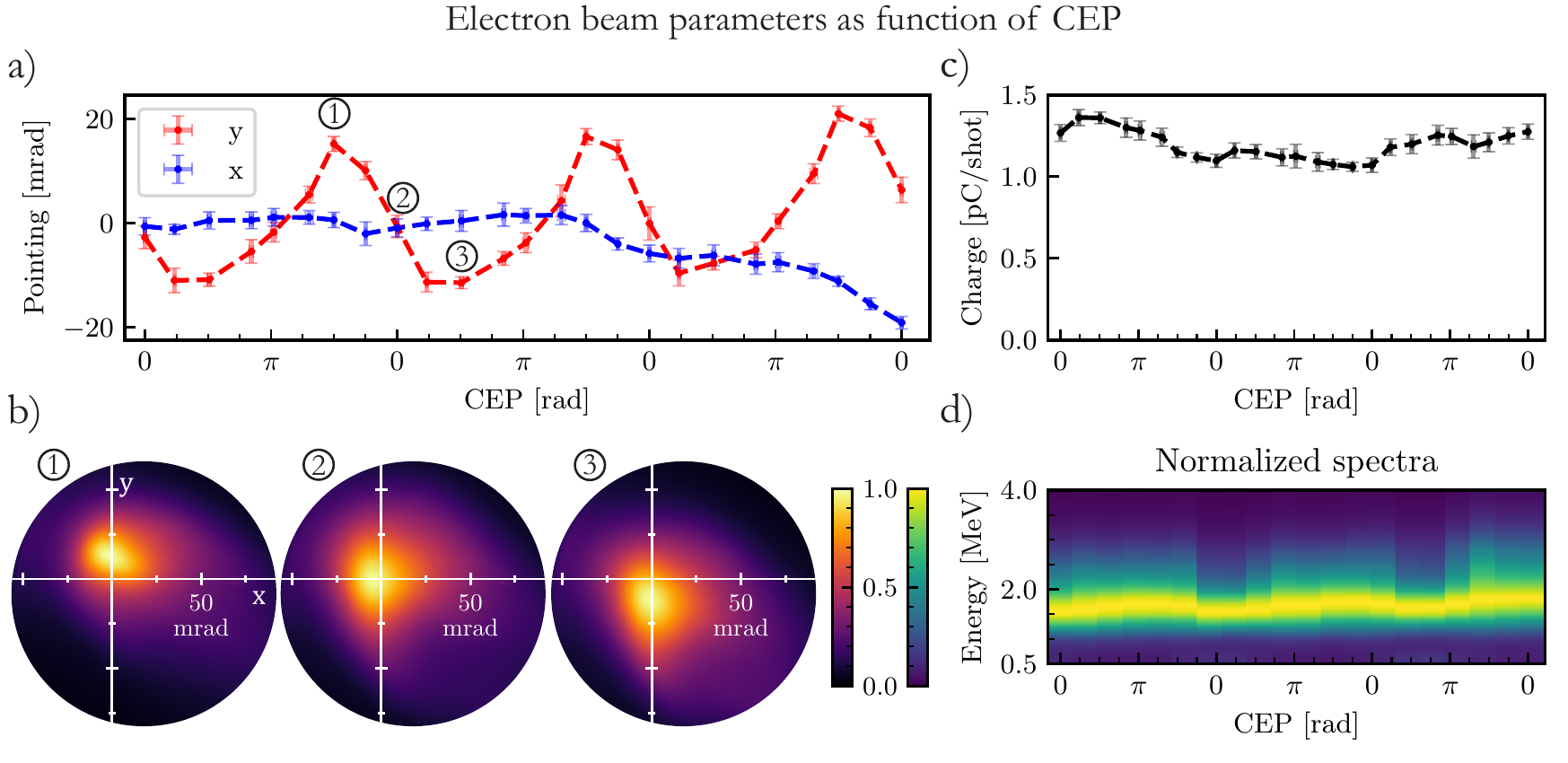}
    \caption{\label{fig:experimental}\textbf{Experimental results} showing changes in electron beam parameters as the CEP is varied over three cycles of $2\pi$. \textbf{a)} The pointing of the electron beam in the plane of polarization ($y$, red) and in the perpendicular plane ($x$, blue). \textbf{b)} Typical images of the electron beam (acquired in 200 ms which corresponds to 200 shots) at a high (1), central (2), and low (3) beam pointing. \textbf{c)} Evolution of the electron beam charge as a function of the CEP. \textbf{d)} The normalized energy spectra of the electron beam as a function of the CEP. Each data point in a) and b) is the average of 20 acquisitions. The vertical error bars indicate the RMS error of these acquisitions, yielding a sub-2 mrad pointing jitter (RMS). The horizontal error bars indicate the RMS error of the CEP stability, averaged over 200 shots (on the order of 40 mrad).}
\end{figure*}

As we vary the CEP, the pointing of the electron beam varies accordingly, as seen in figure~\ref{fig:experimental}a-b (and Supplementary Video 1). This effect is very significant: the amplitude of the oscillation is about 15 mrad, for a beam divergence of around 50 mrad, i.e. a $\sim 30\%$ change. The pointing varies in the plane of the laser polarization ($y$), while in the perpendicular plane ($x$) the beam pointing is constant except for a slow drift. Pointing jitter is below 2 mrad RMS. The beam charge is in the pC range and does not appear to be correlated to CEP changes, see figure~\ref{fig:experimental}c. The electron energy spectrum oscillates in phase with the CEP, albeit with a moderate amplitude of 10 \% of its peak value of 1.62 MeV, see figure~\ref{fig:experimental}d.  This spectral oscillation is due to a combination of two different effects: first, a direct effect of the CEP, i.e. a different CEP yields a different energy distribution. Second, an indirect effect: as the beam pointing changes with the CEP, the electron beam position at the entrance of the spectrometer is different and a different part of the beam is sampled, leading to variations in the measured spectrum. This interpretation is supported by our simulation results (see Supplementary Material for a detailed analysis).

\begin{figure*}
    \includegraphics[width=.8\textwidth]{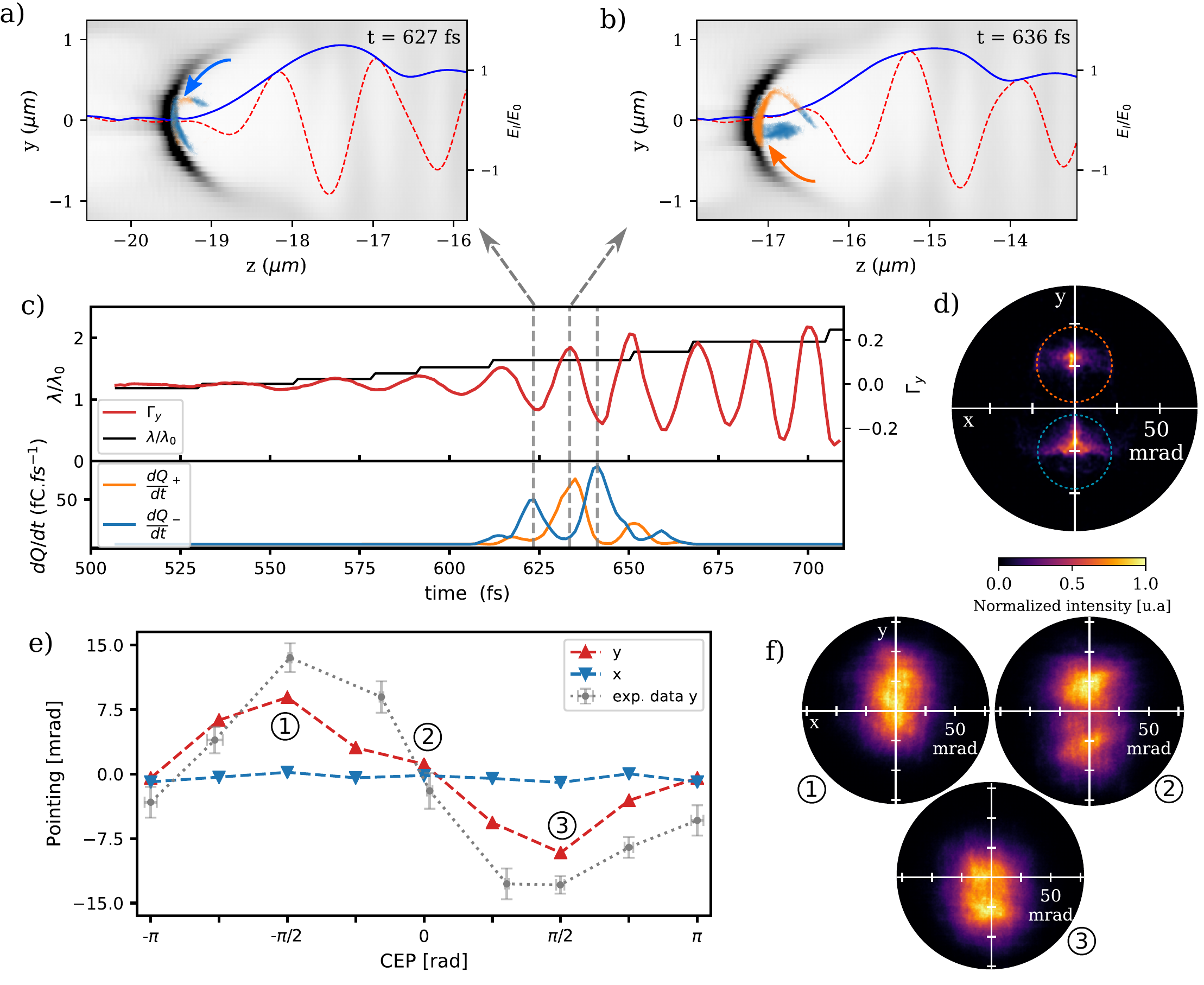}
    \caption{\label{fig:simu} \textbf{Results of PIC simulations.} \textbf{a)-b)} Snapshots of the wakefield, for an initial CEP of $\pi$, at two different times, showing the injection of two separate bunches. Electron density in the z-y plane is showed in gray, injected electrons are displayed in orange (blue) when their pointing is positive (negative) at the end of the simulation. The normalized laser electric field $E_l=m_ec\omega_0/e$  (red dashed line) and its envelope (blue solid line) are also shown. The arrows show the typical trajectories prior to injection for each case.  \textbf{c)} Wakefield oscillation in the polarization plane (red), peak wavelength of the laser normalized by the initial wavelength (black) and charge injection rate for the two electron populations shown in a)-b) with corresponding colors, as a function of the simulation time for an initial CEP of $\pi$. The gray dashed lines highlight the three main injection events.   \textbf{d)} Simulated electron beam for an initial CEP of $\pi$ (single shot). \textbf{e)} Electron beam pointing in the simulations in the directions parallel (red) and transverse (blue) to the laser polarization as a function of the initial CEP. The corresponding experimental data is shown in gray.  \textbf{f)} Simulated electron beams for initial CEP values of $-\frac{\pi}{2}$, 0, $+\frac{\pi}{2}$, which produce a high, centred and low beam, respectively. The experimental beam pointing jitter is simulated by averaging 200 simulated shots with randomly generated beam pointing variations following a normal distribution with a 15\,mrad standard deviation (shot-to-shot) in both $x$ and $y$.}
\end{figure*}

Particle-in-cell simulations using a spectral, quasi-cylindrical particle-in-cell code (see Methods) were carried out in order to reproduce the experimental results and gain more insight into the effect of the CEP on the injection and acceleration process. Typical simulated electron beams have a charge around 2.7\,pC, and originate almost exclusively from self-injection (95\%), with a mean energy of 4.3\,MeV. The difference between the experimental and simulated beams parameters can be explained by the fact that the simulation was run with an idealized case with Gaussian temporal and spatial profiles in order to focus on clarifying the underlying physical process. Figure~\ref{fig:simu}a shows a first injection event that occurs off-axis in the asymmetric wakefield, in the laser polarization plane. As the CEP slips by $\pi$, a second injection event occurs on the other side of the wakefield (panel b). These two electron bunches are injected with opposite initial transverse momenta and they end up with an opposite pointing when they exit the plasma, figure~\ref{fig:simu}d. 

For a more quantitative analysis, we define the wake asymmetry using the electron density transverse centroid, normalized to the the laser waist $w_0$ : $\Gamma_y = \frac{\int n_e y dy}{w_0\int n_e dy}$. In figure~\ref{fig:simu}c, we first confirm that the wake oscillates transversely following the slippage of the CEP (red line). As the laser propagates, its interaction with the wakefield causes a strong redshift\cite{tsun02} (black line). This redshift enhances the wakefield asymmetry in the direction of polarization, following the scaling\cite{huijts2021identifying} $\Gamma_y\propto \lambda^3$.  When $\Gamma_y$ becomes large enough, it triggers the injection of sub-femtosecond electron bunches each time the oscillation reaches an extremum. Note that the injection is very localized and consists in the injection of four sub-bunches, two of which contain most of the charge. This process clearly shows sub-cycle, waveform control over the electron dynamics: electron injection is controlled by the wakefield asymmetry, which is itself controlled by the CEP.

The simulated beam in figure~\ref{fig:simu}d corresponds to a single shot, while in the experiment the data is accumulated over 200 shots and averages over a shot-to-shot pointing jitter that we estimate at 15 mrad RMS. To account for these fluctuations, we numerically added this jitter to the simulated data so that the two sub-bunches merge into a single larger divergence beam, so as to emulate the experimental behaviour, see figure~\ref{fig:simu}f. We were also able to reproduce the dependence of the beam pointing with the CEP: the simulated beam oscillates by 9\,mrad which is comparable to the 15\,mrad obtained in the experiment, see figure~\ref{fig:simu}e-f. The overall change of the beam pointing with the CEP can be understood in the following way: the CEP determines the initial condition of the injected beams, namely the initial longitudinal position in the plasma and initial transverse momentum $z_0,p_{y0}$. The dynamics of the electron beam are then completely deterministic and consist of a transverse oscillation in the wakefield followed by several transverse oscillations in the laser field (see Supplementary Video 1). The final beam pointing is therefore completely determined by the CEP-controlled initial conditions.

These results present experimental evidence that the injection of electrons and the beam parameters are governed directly by the laser waveform, demonstrating the importance of controlling the absolute phase of the field. While in the present experiment electrons are injected in several electron bunches, a future goal is to restrict injection to a single sub-femtosecond bunch. Injection schemes based on ionization will help in achieving this as ionization depends even more directly on the amplitude of the electric field\cite{lifschitz_optical_2012}. Finally, sub-cycle control of electron injection might be very relevant to plasma wakefield accelerators\cite{blum07} driven by particle beams. In this context, sub-cycle, waveform controlled injection with a single cycle laser pulse could provide attosecond and low emittance ultra-relativistic electron beams\cite{hidd2012} for X-ray production.

\section*{Methods}
\subsection*{Laser system and CEP control}
The experiment was conducted using a double Chirped-Pulse Amplification (CPA) system delivering 10\,mJ, 25\,fs pulses at a center wavelength of 800\,nm at a repetition rate of 1\,kHz. The laser spectrum is then broadened by propagating in a 2.5\,m long hollow-core fiber differentially filled with helium \cite{bohle2014,ouille_relativistic-intensity_2020} and re-compressed with chirped mirrors to 4\,fs pulses centred at $\lambda_0$ = 760\,nm. The pulse duration is measured in vacuum using the d-scan technique\cite{miranda12}.  The final energy on target is 2.5\,mJ per pulse. The laser beam is tightly focused by a f/2 off-axis parabola to a $2.7\times2.8\,\mathrm{\mu m}$ FWHM spot, resulting in a vacuum peak intensity of I = $5\times10^{18}\,\mathrm{W.cm^{-2}}$. 
The CEP stabilization is done in two loops, the first of which is a fast feedback loop on the oscillator which modulates the power of the pump laser (managed by an XPS800 by Menlo Systems, Garching, Germany). A second feedback loop stabilizes the CEP after amplification, spectral broadening and compression. It uses a wedge reflection of our probe beam (a fraction of the main beam split off by a holed mirror and used for plasma diagnostics) which is sent to an f-2f interferometer \cite{kakehata2001single} which consists of a \textbeta-barium borate crystal for frequency doubling and a polarizer to project the fundamental and second harmonic polarizations onto the same axis. The interference spectrum is analyzed shot-to-shot by a Fringeezz \cite{lucking2014approaching} (Fastlite, Antibes, France) to measure changes in the CEP. This measurement is fed back to an acousto-optic programmable dispersive filter (Fastlite, Antibes, France) in the first amplification stage to stabilize the CEP. This system stabilizes the CEP with a shot-to-shot dispersion between 240\,mrad RMS and 550\,mrad RMS depending on the target value (see Supplementary Material). As the system measures CEP changes, and not the absolute CEP, there is an arbitrary offset between the curves from the experiment and from the simulation. We shifted the simulated curve in figure \ref{fig:simu}e such that the two curves are in phase.

\subsection*{Target and detectors}
The laser is focused 150\,$\mathrm{\mu m}$ from the exit of a supersonic ``de Laval" nozzle with a 60\,$\mathrm{\mu m}$ throat, and 180\,$\mathrm{\mu m}$ exit diameter. The gas jet flows continuously thanks to a pumping system which keeps the residual gas pressure inside the chamber below $10^{-2}$\,mbar. We measured the gas jet density profile using a quadri-wave lateral shearing interferometer and deduced the plasma density assuming full L-shell ionization of neutral nitrogen $N_2$ into $N^{5+}$. We estimate the peak plasma density obtained in the experiment with a 15\,bar backing pressure to be about $n_e=1.4\times10^{20}\,\mathrm{cm^{-3}}$. In order to keep the density profile as constant as possible throughout the experiment, a pressure controller ensures a sub-percent stability on the backing pressure applied to the gas jet. The electron beam charge and distribution are measured with a calibrated CsI(Tl) phosphor screen imaged on a CCD camera. The energy of the electrons is measured by inserting a removable spectrometer consisting of a 500\,$\mathrm{\mu m}$ pinhole and two permanent circular magnets providing a 58\,mT magnetic field. During the experiments, the continuously flowing gas jet allows us to operate the laser-plasma accelerator at the actual repetition rate of 1\,kHz.

\subsection*{Simulations}
For the simulations we have used a fully relativistic electromagnetic Particle-in-Cell code FBPIC\cite{lehe_spectral_2016} equipped with the pseudo-spectral analytical time domain (PSATD) quasi-cylindrical solver. The PSATD electromagnetic solver is free of numerical dispersion and it provides high-accuracy description of laser propagation and laser-particles interactions, and quasi-cylindrical geometry allows to obtain correct three-dimensional description at a moderate computational cost. The mesh used for simulations is $\Delta z$ = $\lambda_0/60$ and $\Delta r=5\Delta z$. Five azimuthal Fourier modes were used to properly capture all asymmetries. The simulations were initialized with pure neutral Nitrogen, and ionization was calculated with the ADK model of tunnel ionization \cite{ammosov_tunnel_1986}. Atomic nitrogen was initialized using 96 macroparticles per r-z cell, and each such macroparticle could produce up to 7 macroparticles of electron species via ionization.

Idealized Gaussian temporal and spatial laser profiles were used, with waist and pulse duration matching the experiment, and a pulse energy of 2.3\,mJ. Dispersion in the plasma was pre-compensated by adding a $5\,\mathrm{fs}^2$ positive chirp. For the simulated plasma profile, we used a combination of two supergaussian functions to fit the experimentally measured profile, with a peak density of $1.8\times10^{20}\mathrm{cm^{-3}}$. The laser focus position was placed $\SI{25}{\micro\meter}$ upstream of the center of the profile.\par

\bibliography{biblio} 

\begin{thebibliography}{10}
\expandafter\ifx\csname url\endcsname\relax
  \def\url#1{\texttt{#1}}\fi
\expandafter\ifx\csname urlprefix\endcsname\relax\def\urlprefix{URL }\fi
\providecommand{\bibinfo}[2]{#2}
\providecommand{\eprint}[2][]{\url{#2}}

\bibitem{tajima_laser_1979}
\bibinfo{author}{Tajima, T.} \& \bibinfo{author}{Dawson, J.~M.}
\newblock \bibinfo{title}{Laser {Electron} {Accelerator}}.
\newblock \emph{\bibinfo{journal}{Phys. Rev. Lett.}}
  \textbf{\bibinfo{volume}{43}}, \bibinfo{pages}{267--270}
  (\bibinfo{year}{1979}).

\bibitem{esarey_physics_2009}
\bibinfo{author}{Esarey, E.}, \bibinfo{author}{Schroeder, C.~B.} \&
  \bibinfo{author}{Leemans, W.~P.}
\newblock \bibinfo{title}{Physics of laser-driven plasma-based electron
  accelerators}.
\newblock \emph{\bibinfo{journal}{Rev. Mod. Phys.}}
  \textbf{\bibinfo{volume}{81}}, \bibinfo{pages}{1229--1285}
  (\bibinfo{year}{2009}).

\bibitem{corde_femtosecond_2013}
\bibinfo{author}{Corde, S.} \emph{et~al.}
\newblock \bibinfo{title}{Femtosecond x rays from laser-plasma accelerators}.
\newblock \emph{\bibinfo{journal}{Rev. Mod. Phys.}}
  \textbf{\bibinfo{volume}{85}}, \bibinfo{pages}{1--48} (\bibinfo{year}{2013}).

\bibitem{jones2000carrier}
\bibinfo{author}{Jones, D.~J.} \emph{et~al.}
\newblock \bibinfo{title}{Carrier-envelope phase control of femtosecond
  mode-locked lasers and direct optical frequency synthesis}.
\newblock \emph{\bibinfo{journal}{Science}} \textbf{\bibinfo{volume}{288}},
  \bibinfo{pages}{635--639} (\bibinfo{year}{2000}).

\bibitem{diddams2001optical}
\bibinfo{author}{Diddams, S.~A.} \emph{et~al.}
\newblock \bibinfo{title}{An optical clock based on a single trapped 199hg+
  ion}.
\newblock \emph{\bibinfo{journal}{Science}} \textbf{\bibinfo{volume}{293}},
  \bibinfo{pages}{825--828} (\bibinfo{year}{2001}).

\bibitem{paulus_absolute_2001}
\bibinfo{author}{Paulus, G.} \emph{et~al.}
\newblock \bibinfo{title}{Absolute-phase phenomena in photoionization with
  few-cycle laser pulses}.
\newblock \emph{\bibinfo{journal}{Nature}} \textbf{\bibinfo{volume}{414}},
  \bibinfo{pages}{182--184} (\bibinfo{year}{2001}).

\bibitem{baltuska_attosecond_2003}
\bibinfo{author}{Baltu{\v s}ka, A.} \emph{et~al.}
\newblock \bibinfo{title}{Attosecond control of electronic processes by intense
  light fields}.
\newblock \emph{\bibinfo{journal}{Nature}} \textbf{\bibinfo{volume}{421}},
  \bibinfo{pages}{611--615} (\bibinfo{year}{2003}).

\bibitem{nisoli2003effects}
\bibinfo{author}{Nisoli, M.} \emph{et~al.}
\newblock \bibinfo{title}{Effects of carrier-envelope phase differences of
  few-optical-cycle light pulses in single-shot high-order-harmonic spectra}.
\newblock \emph{\bibinfo{journal}{Phys. Rev. Lett.}}
  \textbf{\bibinfo{volume}{91}}, \bibinfo{pages}{213905}
  (\bibinfo{year}{2003}).

\bibitem{kling_control_2006}
\bibinfo{author}{Kling, M.} \emph{et~al.}
\newblock \bibinfo{title}{Control of electron localization in molecular
  dissociation}.
\newblock \emph{\bibinfo{journal}{Science}} \textbf{\bibinfo{volume}{312}},
  \bibinfo{pages}{246--248} (\bibinfo{year}{2006}).

\bibitem{borot2012attosecond}
\bibinfo{author}{Borot, A.} \emph{et~al.}
\newblock \bibinfo{title}{Attosecond control of collective electron motion in
  plasmas}.
\newblock \emph{\bibinfo{journal}{Nature Phys.}} \textbf{\bibinfo{volume}{8}},
  \bibinfo{pages}{416--421} (\bibinfo{year}{2012}).

\bibitem{wheeler2012attosecond}
\bibinfo{author}{Wheeler, J.~A.} \emph{et~al.}
\newblock \bibinfo{title}{Attosecond lighthouses from plasma mirrors}.
\newblock \emph{\bibinfo{journal}{Nature Photon.}}
  \textbf{\bibinfo{volume}{6}}, \bibinfo{pages}{829--833}
  (\bibinfo{year}{2012}).

\bibitem{kormin2018spectral}
\bibinfo{author}{Kormin, D.} \emph{et~al.}
\newblock \bibinfo{title}{Spectral interferometry with waveform-dependent
  relativistic high-order harmonics from plasma surfaces}.
\newblock \emph{\bibinfo{journal}{Nat. Commun.}} \textbf{\bibinfo{volume}{9}},
  \bibinfo{pages}{1--8} (\bibinfo{year}{2018}).

\bibitem{jahn2019towards}
\bibinfo{author}{Jahn, O.} \emph{et~al.}
\newblock \bibinfo{title}{Towards intense isolated attosecond pulses from
  relativistic surface high harmonics}.
\newblock \emph{\bibinfo{journal}{Optica}} \textbf{\bibinfo{volume}{6}},
  \bibinfo{pages}{280--287} (\bibinfo{year}{2019}).

\bibitem{bohle2020generation}
\bibinfo{author}{B{\"o}hle, F.} \emph{et~al.}
\newblock \bibinfo{title}{Generation of {XUV} spectral continua from
  relativistic plasma mirrors driven in the near-single-cycle limit}.
\newblock \emph{\bibinfo{journal}{J. Phys. Photonics}}
  \textbf{\bibinfo{volume}{2}}, \bibinfo{pages}{034010} (\bibinfo{year}{2020}).

\bibitem{nerush_carrier-envelope_2009}
\bibinfo{author}{Nerush, E.~N.} \& \bibinfo{author}{Kostyukov, I.~Y.}
\newblock \bibinfo{title}{Carrier-{Envelope} {Phase} {Effects} in
  {Plasma}-{Based} {Electron} {Acceleration} with {Few}-{Cycle} {Laser}
  {Pulses}}.
\newblock \emph{\bibinfo{journal}{Phys. Rev. Lett.}}
  \textbf{\bibinfo{volume}{103}}, \bibinfo{pages}{035001}
  (\bibinfo{year}{2009}).

\bibitem{faure_review_2018}
\bibinfo{author}{Faure, J.} \emph{et~al.}
\newblock \bibinfo{title}{A review of recent progress on laser-plasma
  acceleration at {kHz} repetition rate}.
\newblock \emph{\bibinfo{journal}{Plasma Phys. Control. Fusion}}
  \textbf{\bibinfo{volume}{61}}, \bibinfo{pages}{014012}
  (\bibinfo{year}{2018}).

\bibitem{geddes_high-quality_2004}
\bibinfo{author}{Geddes, C. G.~R.} \emph{et~al.}
\newblock \bibinfo{title}{High-quality electron beams from a laser wakefield
  accelerator using plasma-channel guiding}.
\newblock \emph{\bibinfo{journal}{Nature}} \textbf{\bibinfo{volume}{431}},
  \bibinfo{pages}{538--541} (\bibinfo{year}{2004}).

\bibitem{faure_laserplasma_2004}
\bibinfo{author}{Faure, J.} \emph{et~al.}
\newblock \bibinfo{title}{A laser--plasma accelerator producing monoenergetic
  electron beams}.
\newblock \emph{\bibinfo{journal}{Nature}} \textbf{\bibinfo{volume}{431}},
  \bibinfo{pages}{541--544} (\bibinfo{year}{2004}).

\bibitem{mangles_monoenergetic_2004}
\bibinfo{author}{Mangles, S. P.~D.} \emph{et~al.}
\newblock \bibinfo{title}{Monoenergetic beams of relativistic electrons from
  intense laser--plasma interactions}.
\newblock \emph{\bibinfo{journal}{Nature}} \textbf{\bibinfo{volume}{431}},
  \bibinfo{pages}{535--538} (\bibinfo{year}{2004}).

\bibitem{schmid_few-cycle_2009}
\bibinfo{author}{Schmid, K.} \emph{et~al.}
\newblock \bibinfo{title}{Few-{Cycle} {Laser}-{Driven} {Electron}
  {Acceleration}}.
\newblock \emph{\bibinfo{journal}{Phys. Rev. Lett.}}
  \textbf{\bibinfo{volume}{102}}, \bibinfo{pages}{124801}
  (\bibinfo{year}{2009}).

\bibitem{guenot_relativistic_2017}
\bibinfo{author}{Gu{\'e}not, D.} \emph{et~al.}
\newblock \bibinfo{title}{Relativistic electron beams driven by {kHz}
  single-cycle light pulses}.
\newblock \emph{\bibinfo{journal}{Nature Photon.}}
  \textbf{\bibinfo{volume}{11}}, \bibinfo{pages}{293--296}
  (\bibinfo{year}{2017}).

\bibitem{rovige_optimization_2021}
\bibinfo{author}{Rovige, L.} \emph{et~al.}
\newblock \bibinfo{title}{Optimization and stabilization of a kilohertz
  laser-plasma accelerator}.
\newblock \emph{\bibinfo{journal}{Physics of Plasmas}}
  \textbf{\bibinfo{volume}{28}}, \bibinfo{pages}{033105}
  (\bibinfo{year}{2021}).

\bibitem{salehi2020laseraccelerated}
\bibinfo{author}{Salehi, F.}, \bibinfo{author}{Le, M.},
  \bibinfo{author}{Railing, L.} \& \bibinfo{author}{Milchberg, H.~M.}
\newblock \bibinfo{title}{Laser-accelerated, low divergence {15~MeV}
  quasi-monoenergetic electron bunches at {1~kHz}} (\bibinfo{year}{2020}).
\newblock arXiv:\eprint{2010.15720}.

\bibitem{mora_kinetic_1997}
\bibinfo{author}{Mora, P.} \& \bibinfo{author}{Antonsen, T.~M., Jr}.
\newblock \bibinfo{title}{Kinetic modeling of intense, short laser pulses
  propagating in tenuous plasmas}.
\newblock \emph{\bibinfo{journal}{Physics of Plasmas}}
  \textbf{\bibinfo{volume}{4}}, \bibinfo{pages}{217--229}
  (\bibinfo{year}{1997}).

\bibitem{zhidkov_electron_2008}
\bibinfo{author}{Zhidkov, A.}, \bibinfo{author}{Fujii, T.} \&
  \bibinfo{author}{Nemoto, K.}
\newblock \bibinfo{title}{Electron self-injection during interaction of tightly
  focused few-cycle laser pulses with underdense plasma}.
\newblock \emph{\bibinfo{journal}{Phys. Rev. E}} \textbf{\bibinfo{volume}{78}},
  \bibinfo{pages}{036406} (\bibinfo{year}{2008}).

\bibitem{huijts2021identifying}
\bibinfo{author}{Huijts, J.}, \bibinfo{author}{Andriyash, I.~A.},
  \bibinfo{author}{Rovige, L.}, \bibinfo{author}{Vernier, A.} \&
  \bibinfo{author}{Faure, J.}
\newblock \bibinfo{title}{Identifying observable carrier-envelope phase effects
  in laser wakefield acceleration with near-single-cycle pulses}.
\newblock \emph{\bibinfo{journal}{Physics of Plasmas}}
  \textbf{\bibinfo{volume}{28}}, \bibinfo{pages}{043101}
  (\bibinfo{year}{2021}).

\bibitem{xu_periodic_2020}
\bibinfo{author}{Xu, S.} \emph{et~al.}
\newblock \bibinfo{title}{Periodic self-injection of electrons in a few-cycle
  laser driven oscillating plasma wake}.
\newblock \emph{\bibinfo{journal}{AIP Advances}} \textbf{\bibinfo{volume}{10}},
  \bibinfo{pages}{095310} (\bibinfo{year}{2020}).

\bibitem{tsun02}
\bibinfo{author}{Tsung, F.~S.}, \bibinfo{author}{Ren, C.},
  \bibinfo{author}{Silva, L.~O.}, \bibinfo{author}{Mori, W.~B.} \&
  \bibinfo{author}{Katsouleas, T.}
\newblock \bibinfo{title}{Generation of ultra-intense single-cycle laser pulses
  by using photon deceleration}.
\newblock \emph{\bibinfo{journal}{Proc. Nat. Acad. Science}}
  \textbf{\bibinfo{volume}{99}}, \bibinfo{pages}{29--32}
  (\bibinfo{year}{2002}).

\bibitem{lifschitz_optical_2012}
\bibinfo{author}{Lifschitz, A.~F.} \& \bibinfo{author}{Malka, V.}
\newblock \bibinfo{title}{Optical phase effects in electron wakefield
  acceleration using few-cycle laser pulses}.
\newblock \emph{\bibinfo{journal}{New J. Phys.}} \textbf{\bibinfo{volume}{14}},
  \bibinfo{pages}{053045} (\bibinfo{year}{2012}).

\bibitem{blum07}
\bibinfo{author}{Blumenfeld, I.} \emph{et~al.}
\newblock \bibinfo{title}{Energy doubling of 42 {GeV} electrons in a
  metre-scale plasma wakefield accelerator}.
\newblock \emph{\bibinfo{journal}{Nature}}  (\bibinfo{year}{2007}).

\bibitem{hidd2012}
\bibinfo{author}{Hidding, B.} \emph{et~al.}
\newblock \bibinfo{title}{Ultracold electron bunch generation via plasma
  photocathode emission and acceleration in a beam-driven plasma blowout}.
\newblock \emph{\bibinfo{journal}{Phys. Rev. Lett.}}
  \textbf{\bibinfo{volume}{108}}, \bibinfo{pages}{035001}
  (\bibinfo{year}{2012}).

\bibitem{bohle2014}
\bibinfo{author}{B{\"o}hle, F.} \emph{et~al.}
\newblock \bibinfo{title}{Compression of {CEP}-stable multi-{mJ} laser pulses
  down to 4{\hspace{0.167em}}fs in long hollow fibers}.
\newblock \emph{\bibinfo{journal}{Laser Phys. Lett.}}
  \textbf{\bibinfo{volume}{11}}, \bibinfo{pages}{095401}
  (\bibinfo{year}{2014}).

\bibitem{ouille_relativistic-intensity_2020}
\bibinfo{author}{Ouill{\'e}, M.} \emph{et~al.}
\newblock \bibinfo{title}{Relativistic-intensity near-single-cycle light
  waveforms at {kHz} repetition rate}.
\newblock \emph{\bibinfo{journal}{Light. Sci. Appl.}}
  \textbf{\bibinfo{volume}{9}}, \bibinfo{pages}{1--9} (\bibinfo{year}{2020}).

\bibitem{miranda12}
\bibinfo{author}{Miranda, M.} \emph{et~al.}
\newblock \bibinfo{title}{Characterization of broadband few-cycle laser pulses
  with the d-scan technique}.
\newblock \emph{\bibinfo{journal}{Opt. Express}} \textbf{\bibinfo{volume}{20}},
  \bibinfo{pages}{18732--18743} (\bibinfo{year}{2012}).

\bibitem{kakehata2001single}
\bibinfo{author}{Kakehata, M.} \emph{et~al.}
\newblock \bibinfo{title}{Single-shot measurement of carrier-envelope phase
  changes by spectral interferometry}.
\newblock \emph{\bibinfo{journal}{Opt. Lett.}} \textbf{\bibinfo{volume}{26}},
  \bibinfo{pages}{1436--1438} (\bibinfo{year}{2001}).

\bibitem{lucking2014approaching}
\bibinfo{author}{L{\"u}cking, F.}, \bibinfo{author}{Crozatier, V.},
  \bibinfo{author}{Forget, N.}, \bibinfo{author}{Assion, A.} \&
  \bibinfo{author}{Krausz, F.}
\newblock \bibinfo{title}{Approaching the limits of carrier-envelope phase
  stability in a millijoule-class amplifier}.
\newblock \emph{\bibinfo{journal}{Opt. Lett.}} \textbf{\bibinfo{volume}{39}},
  \bibinfo{pages}{3884--3887} (\bibinfo{year}{2014}).

\bibitem{lehe_spectral_2016}
\bibinfo{author}{Lehe, R.}, \bibinfo{author}{Kirchen, M.},
  \bibinfo{author}{Andriyash, I.~A.}, \bibinfo{author}{Godfrey, B.~B.} \&
  \bibinfo{author}{Vay, J.-L.}
\newblock \bibinfo{title}{A spectral, quasi-cylindrical and dispersion-free
  {Particle}-{In}-{Cell} algorithm}.
\newblock \emph{\bibinfo{journal}{Comput. Phys. Commun.}}
  \textbf{\bibinfo{volume}{203}}, \bibinfo{pages}{66--82}
  (\bibinfo{year}{2016}).

\bibitem{ammosov_tunnel_1986}
\bibinfo{author}{Ammosov, M.}, \bibinfo{author}{Delone, N.} \&
  \bibinfo{author}{Krainov, V.}
\newblock \bibinfo{title}{Tunnel ionization of complex atoms and of atomic ions
  in an alternating electromagnetic field}.
\newblock \emph{\bibinfo{journal}{Sov. Phys. JETP}}
  \textbf{\bibinfo{volume}{96}} (\bibinfo{year}{1986}).

\end{thebibliography}

\begin{acknowledgements}
This work was funded by the Agence Nationale de la Recherche under Contract Numbers. ANR-20-CE92-0043-01. Financial support from the European Research Council (ERC Starting Grant FEMTOELEC 306708, ERC Advanced Grant ExCoMet 694596) is gratefully acknowledged. Numerical simulations presented in this work were done using resources of TGCC HPC cluster under the allocation of GENCI project A0090510062.
\end{acknowledgements}

\section*{Author contributions} L.R. and J.H. performed the experiment. J.H. analyzed the experimental data. L.R and I.A. performed the simulations. A.V. built the experimental apparatus and participated also in the experiment. M.O., J.K., Z.C. and R.LM. developed and took care of the laser system and implemented the control of the CEP. J.F. directed the project and participated in the interpretation of the data. The paper was written by L.R., J.H. and J.F. with the inputs from all other authors.

\section*{Competing Interests}
The authors declare no competing interests.

\newpage

\newpage

\end{document}